\documentclass[a4paper]{jpconf}
\usepackage{graphicx}
\usepackage{amsmath, amssymb}
\usepackage{caption}
\usepackage{multirow}
\usepackage{tabularx}
\usepackage{bbm}
\usepackage[usenames, dvipsnames]{color}
\begin{document}
\begin{flushright}
INR-TH-2017-020
\end{flushright}
\title{ The analytical $\mathcal{O}(a^4_s)$ expression for the polarized Bjorken sum rule in the miniMOM scheme and the  consequences for the generalized Crewther 
relation}

\author{ A. L. Kataev$^{1,2}$, V. S. Molokoedov$^{1,2,3}$ }

\address{$^1$ Institute for Nuclear Research of the Russian Academy of Sciences (INR), 60th October Anniversary Prospect, 7a, 117312 Moscow, Russia }
\address{$^2$ Moscow Institute of Physics and Technology (MIPT), 141700, Dolgoprudnyy, Russia}
\address{$^3$ Landau Institute for Theoretical Physics of the Academy of Sciences of Russia, 142432, Moscow Region, Russia}

\ead{kataev@ms2.inr.ac.ru, viktor\_molokoedov@mail.ru}

\begin{abstract}
The analytical  $\mathcal{O}(a^4_s)$ perturbative QCD 
 expression for the flavour non-singlet contribution to 
the Bjorken polarized sum rule in the rather  applicable at present  
gauge--dependent $\rm{miniMOM}$ scheme is obtained. 
For the considered three values of the gauge parameter, namely 
$\xi=0$  (Landau gauge), $\xi=-1$ (anti--Feynman gauge) and $\xi=-3$ 
(Stefanis--Mikhailov gauge),  the scheme-dependent coefficients are 
considerably smaller than the gauge-independent  $\rm{\overline{MS}}$ results.
It is found that the fundamental  property of the factorization of the 
QCD renormalization
group $\beta$-function in the generalized Crewther relation, which is 
valid in the gauge-invariant  $\rm{\overline{MS}}$ scheme up to   
$\mathcal{O}(a^4_s)$-level at least, is unexpectedly  valid at the same level 
in the  $\rm{miniMOM}$-scheme for $\xi=0$, and for $\xi=-1$ and $\xi=-3$ 
in part.   
\end{abstract}

\section{Introduction}
It is known that the hadronic tensor, which enters into the definitions 
of differential cross-sections of  deep inelastic scattering (DIS) 
processes of 
charged leptons on nucleons, contains the terms, that are measured in  the  
DIS of  polarized leptons on 
polarized nucleons,
namely 
\begin{eqnarray}
W_{\mu\nu}(p, q, s)&=&\frac{1}{4\pi}\int d^4z \; e^{iqz}\langle p, s\vert [ J^{\dagger}_\mu(z), J_\nu(0)]\vert p, s\rangle \\ \nonumber
&=& \bigg(\frac{q_\mu q_\nu}{q^2}-g_{\mu\nu}\bigg)F_1(x, Q^2) +\frac{1}{(pq)}\bigg(p_\mu-\frac{(pq)}{q^2}q_\mu\bigg)\bigg(p_\nu-\frac{(pq)}{q^2}q_\nu\bigg)F_2(x, Q^2) \\ \nonumber
&+&i\varepsilon_{\mu\nu\lambda\rho}\frac{q^\lambda}{(pq)}\bigg(s^\rho (g_1(x, Q^2)+g_2(x, Q^2))-\frac{(sq)}{(pq)}p^\rho g_2(x, Q^2)\bigg)+\dots ~, 
\end{eqnarray}
where $J^{\mu}(z)=\sum_f Q_f \bar{q}_f(z)\gamma^{\mu}q_f(z)$ is the 
quark electromagnetic current, $p$ is the four--momentum of nucleon and $s$ 
is its spin, $q$ is the transferred momentum with $Q^2=-q^2$, $0\leq x=Q^2/(2pq)\leq 1$ is the Bjorken  variable. In this expression the structure functions  
$g_1(x,Q^2)$ and $g_2(x,Q^2)$ are the ones which are extracted from 
differential cross-section of DIS process of polarized charged leptons on
polarized nucleons, while    $F_1(x, Q^2)$ and 
$F_2(x, Q^2)$  characterize the DIS processes with  unpolarized particles.

The important characteristic of DIS  process  of polarized charged 
leptons on polarized nucleons  is the Bjorken polarized sum rule, defined as  
\begin{equation} 
\label{Bjp}
\int\limits_0^1 \bigg(g^{lp}_1(x, Q^2)-g^{ln}_1(x, Q^2)\bigg)dx =
\frac{1}{6}\bigg\vert\frac{g_A}{g_V}\bigg\vert
C_{Bjp}(Q^2)~,
\end{equation}
where index $l$ defines the polarized charged lepton ($e$  or $\mu$) in 
the concrete different experiments. The structure functions 
$g_1^{lp(n)}(x, Q^2)$ characterize the spin distribution of polarized quarks and gluons 
inside nucleons. The world 
average value of the ratio of   axial and vector  charges  of 
neutron $\beta$ decay is  $g_A/g_V=-1.2723 \pm 0.0023$  
\cite{Patrignani:2016xqp}. 
The general theoretical expression 
for the  Bjorken polarized  sum rule contains massive-dependent corrections 
\cite{Teryaev:1996nf}, \cite{Blumlein:1998sh} 
and non-perturbative high-twist  $\mathcal{O}(1/Q^2)$--corrections, discussed  in \cite{Kataev:2005hv}. However, in this work we will consider the massless 
PT expression for the coefficient 
function $C_{Bjp}(Q^2)$ in the  $SU(N_c)$ colour group only, 
which is known at the  $\mathcal{O}(a^4_s)$ level in 
the $\rm{\overline{MS}}$-scheme from the results of \cite{Baikov:2010je}, 
supplemented  the singlet contribution, which appears first at the 
same level  \cite{Larin:2013yba} with the numerically small analytical 
coefficient, evaluated in \cite{Baikov:2015tea}. Therefore, in general the 
massless PT expression for $C_{Bjp}(Q^2)$  can be written down as:   
\begin{equation}
\label{BjrNS-SI}
C_{Bjp}(Q^2)=C^{NS}_{Bjp}(Q^2)+\sum_f Q_f C^{SI}_{Bjp}(Q^2)~,
\end{equation}
where 
\begin{equation}   
\label{NS-coef}
C^{NS}_{Bjp}(Q^2)=1+\sum\limits_{n=1}^{\infty} \overline{c}^{\; NS}_n \overline{a}^{\; n}_s(Q^2)~, ~~~~~ 
C^{SI}_{Bjp}(Q^2)=\sum\limits_{n=4}^{\infty} \overline{c}^{\; SI}_n \overline{a}^{\; n}_s(Q^2)~, 
\end{equation}
where $\overline{a}_s=\overline{\alpha}_s/\pi$ is the strong coupling constant determined in $\rm{\overline{MS}}$ scheme.
Remind that the  first, second and third PT contributions to $C^{NS}_{Bjp}(Q^2)$
in the $\rm{\overline{MS}}$-scheme 
were analytically calculated in the works \cite{Kodaira:1978sh}, \cite{Gorishnii:1985xm} and  \cite{Larin:1991tj}
respectively. Note also that for $n_f=3$ the singlet contribution is 
identically equal to zero.   

It is known that  the scheme--dependent  PT series  
for the renormalization--group (RG) 
invariant quantities are asymptotic ones. The 
transformation to  the special schemes, which   minimize 
the contributions of 
high-order PT  corrections to physical quantities,   is one of 
theoretical 
approaches for applying them in  phenomenology. In \cite{Celmaster:1979dm} it was proposed to 
achieve this goal by using the special momentum ($\rm{MOM}$) 
subtractions scheme, 
which in QCD is gauge--dependent. It should be noted, that in the $\rm{MOM}$--like 
schemes the coefficients of the QCD RG  $\beta$-function are 
becoming gauge--dependent starting from the two-loop level 
\cite{Tarasov:1990ps}. The gauge-dependence of the coefficients of both 
RG-invariant  quantities and QCD $\beta$-function in various $\rm{MOM}$ schemes 
makes  the analysis of the PT expansions for the quantities, interesting 
from experimental point of view, more complicated.
However, the interest to the studies of the 
PT approximations for various QCD  characteristics of observable physical 
processes in the special gauge-dependent  $\rm{MOM}$ schemes  became actual 
again. It  is related mainly  to the formulation 
of minimal $\rm{MOM}$ ($\rm{mMOM}$) scheme \cite{vonSmekal:2009ae}. Its 
analog was   
used in studies of the Functional Renormalization Group Equations \cite{Pawlowski:2003hq}. 

In  Section II we remind the main ideas of the  definition of 
$\rm{mMOM}$ scheme and present the explicit analytical gauge--dependent relation 
between the QCD coupling constants in the  $\rm{\overline{MS}}$ and
$\rm{MOM}$ schemes with gauge covariant linear parameter $\xi$, determined in $\rm{mMOM}$ scheme. The  new analytical 
results for the PT $\rm{mMOM}$ coefficients of the  $\mathcal{O}(a^4_s)$ approximation 
of the Bjorken polarized sum rule are presented. Their   
numerical values, 
specified to the Landau gauge  $\xi=0$, to the  anti--Feynman gauge 
 $\xi=-1$ and to the Stefanis--Mikhailov  gauge  $\xi=-3$, first used in QCD 
by Stefanis in the work of \cite{Stefanis:1983ke} (see also \cite{Stefanis:2012if}) to untangle  the special
features of renormalizations of gauge--invariant definition of QCD quark 
correlator, formulated with the help of the Wilson lines,     and independently 
applied later on by Mikhailov in \cite{Mikhailov:1998xi} (see also \cite{Mikhailov:1999ht}) to clarify the 
renormalization of the gluon fields  by  the renormalon chain contributions, 
are compared with the $\rm{\overline{MS}}$ scheme results.     

In Section III we   will clarify  that  the application of the 
anti--Feynman and 
Stefanis--Mikhailov gauges are essential  for satisfying the fundamental  
property of the
factorization of the gauge--dependent  QCD $\beta$-function in 
the $\mathcal{O}(a^3_s)$ approximation in  $\rm{mMOM}$  scheme 
variant of the QCD  generalized  
Crewther relation (GCR),  
 discovered in \cite{Broadhurst:1993ru} in the   
gauge--independent $\rm{\overline{MS}}$ scheme, which in 
the conformal symmetry limit reproduce well-known  massless quark-parton 
result, obtained in  \cite{Crewther:1972kn}.
We will also stress that  
the transformation to the  $\rm{mMOM}$ scheme when the  Landau gauge 
is chosen 
will not spoil the property of 
the factorization of the analytical expression for the QCD  $\beta$-function
in the  
$SU(N_c)$ GCR  
at the fourth  order of PT,   
studied in the   $\rm{\overline{MS}}$ scheme in \cite{Baikov:2010je}.

\section{The Bjorken polarized sum rule in mMOM scheme}

The $\rm{mMOM}$ scheme was introduced in \cite{vonSmekal:2009ae} and was initially formulated as
 the easiest way  to satisfy the common $\rm{MOM}$ schemes property  of the non-renormalization of the gluon--ghost--anti-ghost vertex in the Landau gauge $\xi=0$ thanks to 
the equality between the renormalization constant of this vertex in $\rm{mMOM}$  
and  $\rm{\overline{MS}}$ schemes \cite{vonSmekal:2009ae}: 
\begin{equation}
\label{MOM-MS}
Z^{{\rm{mMOM}}}_{cg}=Z^{\rm{\overline{MS}}}_{cg}~.
\end{equation}    
Taking into account this relation (\ref{MOM-MS}) and the $\rm{MOM}$--like renormalization conditions for gluon and ghost propagators, one can obtain the following relations between strong coupling constant $a^{\rm{mMOM}}_s\equiv a_s$ in $\rm{mMOM}$ scheme and $a^{{\rm{\overline{MS}}}}_s\equiv \overline{a}_s$ and  between gauge parameters $\xi^{\rm{mMOM}}\equiv \xi$ and $\xi^{{\rm{\overline{MS}}}}\equiv\overline{\xi}$ \cite{vonSmekal:2009ae}, \cite{Gracey:2013sca}, \cite{Ruijl:2017eht}: 
\begin{eqnarray}
\label{a-xi}
a_s(\mu^2)=\frac{\overline{a}_s(\mu^2)}{(1+\Delta(\mu^2))(1+\Omega(\mu^2))^2}~, ~~~~~~ 
\xi(\mu^2)=\overline{\xi}(\mu^2)(1+\Delta(\mu^2))~, 
\end{eqnarray}
where  gluon and ghost self-energies functions $\Delta(q^2)$ and 
$\Omega(q^2)$ are evaluated in the  $\rm{\overline{MS}}$ 
scheme and  defined by the higher order corrections to the corresponding propagators. 
In the  arbitrary $SU(N_c)$ colour group their  explicit form was analytically  calculated  in \cite{Chetyrkin:2000dq} at $\mathcal{O}(a^3_s)$ level 
and in \cite{Ruijl:2017eht} in the fourth order PT  approximation. 
Note that these $\rm{\overline{MS}}$ results for self-energies $\Delta(q^2)$ and 
$\Omega(q^2)$ depend on the  gauge parameter starting from one- and two--loop level correspondingly.
The expressions (\ref{a-xi}) allow us to get the following  explicit analytical  three-loop approximation  
of the function  
$\overline{a}_s(a_s,\xi)$: 
\begin{eqnarray}
\label{aMS-mMOM}
\overline{a}_s &=&a_s\bigg(1 +\delta_1 a_s+\delta_2a^2_s+\delta _3 a^3_s+
\mathcal{O}(a^4_s)\bigg)~, \\
\delta _1&=&\bigg(-\frac{169}{144}-\frac{1}{8}{\color{BrickRed}\xi}-\frac{1}{16}{\color{BrickRed}\xi^2}\bigg){\color{blue}
C_A}+\frac{5}{9}{\color{blue} T_Fn_f}~, \\
\delta _2&=&\bigg[-\frac{18941}{20736}+\frac{39}{128}\zeta_3+\bigg(\frac{889}{2304}-\frac{11}{64}\zeta_3\bigg){\color{BrickRed}\xi}+\bigg(\frac{203}{2304}+\frac{3}{128}\zeta_3\bigg){\color{BrickRed}\xi^2}-\frac{3}{256}{\color{BrickRed}\xi^3}\bigg]{\color{blue}
C_A^2} \\ \nonumber
&+&\bigg[-\frac{107}{648}+\frac{\zeta_3}{2}-\frac{5}{36}{\color{BrickRed}\xi}-\frac{5}{72}{\color{BrickRed}\xi^2} \bigg]{\color{blue}
C_AT_Fn_f}+\bigg[\frac{55}{48}-\zeta_3\bigg]{\color{blue} C_FT_Fn_f}+\frac{25}{81}{\color{blue} T^2_Fn^2_f}~, 
\end{eqnarray}
\begin{eqnarray}
\delta _3&=&\bigg[-\frac{1935757}{2985984}+\frac{7495}{18432}\zeta_3+\frac{7805}{12288}\zeta_5+\bigg(\frac{4877}{36864}-\frac{611}{1536}\zeta_3+\frac{295}{1024}\zeta_5\bigg){\color{BrickRed}\xi} \\ \nonumber
&+&\bigg(\frac{17315}{110592}-\frac{47}{768}\zeta_3+\frac{175}{6144}\zeta_5\bigg){\color{BrickRed}\xi^2}+\bigg(-\frac{233}{4608}+\frac{59}{1536}\zeta_3+\frac{5}{3072}\zeta_5\bigg){\color{BrickRed}\xi^3} \\ \nonumber
&+&\bigg(-\frac{235}{36864}-\frac{5}{6144}\zeta_3-\frac{35}{12288}\zeta_5\bigg){\color{BrickRed}\xi^4}\bigg]{\color{blue}C^3_A}+\bigg[-\frac{143}{288}-\frac{37}{24}\zeta_3+\frac{5}{2}\zeta_5\bigg]{\color{blue} C^2_FT_Fn_f} \\ \nonumber
 &+&\bigg[\frac{25547}{20736}+\frac{107}{144}\zeta_3-\frac{5}{4}\zeta_5+\bigg(-\frac{55}{128}+\frac{3}{8}\zeta_3\bigg){\color{BrickRed}\xi}+\bigg(-\frac{55}{256}+\frac{3}{16}\zeta_3\bigg){\color{BrickRed}\xi^2}
 \bigg]{\color{blue} C_FC_AT_Fn_f} \\ \nonumber
 &+&\bigg[\frac{199}{31104}-\frac{223}{384}\zeta_3-\frac{5}{6}\zeta_5+\bigg(\frac{833}{13824}+\frac{11}{144}\zeta_3\bigg){\color{BrickRed}\xi}
+\bigg(-\frac{143}{6912}-\frac{9}{128}\zeta_3\bigg){\color{BrickRed}\xi^2}-\frac{5}{1152}{\color{BrickRed}\xi^3} \\ \nonumber
&+&\frac{5}{2304}{\color{BrickRed}\xi^4}
\bigg]{\color{blue} C^2_AT_Fn_f}+   \bigg[\frac{1235}{15552}+\frac{13}{36}\zeta_3+\bigg(-\frac{19}{216}-\frac{1}{9}\zeta_3\bigg){\color{BrickRed}\xi}
-\frac{25}{432}{\color{BrickRed}\xi^2}\bigg]{\color{blue} C_AT^2_Fn^2_f}
\\ \nonumber
&+&\bigg[\frac{1249}{2592}-\frac{11}{18}\zeta_3\bigg]{\color{blue} C_FT^2_Fn^2_f}+\frac{125}{729}{\color{blue}T^3_Fn^3_f}~,
\end{eqnarray}
where $C_F$ and $C_A$ are the Casimir operators, $T_F$ is the Dynkin index, $n_f$ is the number of active flavours.

In order to obtain values of the coefficients $c^{NS}_n$ of the 
flavour non-singlet Bjorken function in the  $\rm{mMOM}$ scheme we use the 
known four--loop $\rm{\overline{MS}}$ results for $C^{NS}_{Bjp}$ from \cite{Baikov:2010je}
(the  corresponding coefficients are denoted as  $\overline{c}^{NS}_n$),
the  relation (\ref{aMS-mMOM}) and the renorm--invariant property of the physical quantity $C^{NS}_{Bjp}$. 
The relations between the coefficients in $\rm{mMOM}$ and  $\rm{\overline{MS}}$ schemes read: 
\begin{eqnarray}
c^{NS}_1&=&\overline{c}^{\; NS}_1~, ~~~ c^{NS}_2=\overline{c}^{\; NS}_2+\delta_1\overline{c}^{\; NS}_1~, ~~~ c^{NS}_3=\overline{c}^{\; NS}_3+2\delta_1\overline{c}^{\; NS}_2+\delta_2\overline{c}^{\; NS}_1~, \\ 
c^{NS}_4&=&\overline{c}^{\; NS}_4+3\delta_1\overline{c}^{\; NS}_3+(2\delta_2+\delta^2_1)\overline{c}^{\; NS}_2+\delta_3\overline{c}^{\; NS}_1~.
\end{eqnarray}
Using these relations  we find the explicit form of coefficients of the Bjorken function in the  $\rm{mMOM}$ scheme at the  $\mathcal{O}(a^4_s)$ level:
\begin{eqnarray} 
\label{c1mMoM}
c^{NS}_1&=&-\frac{3}{4}{\color{blue} C_F}~, \\ 
\label{c2mMOM}
c^{NS}_2&=&\frac{21}{32}{\color{blue} C^2_F}+\bigg(-\frac{107}{192}+\frac{3}{32}{\color{BrickRed}\xi}+\frac{3}{64}{\color{BrickRed}\xi^2}\bigg){\color{blue} C_FC_A}+\frac{1}{12}{\color{blue} C_FT_Fn_f}~, \\
\label{c3mMOM}
c^{NS}_3&=&-\frac{3}{128}{\color{blue}C^3_F}+
\bigg[\frac{1415}{2304}-\frac{11}{12}\zeta_3 -\frac{21}{128}{\color{BrickRed}\xi}-\frac{21}{256}{\color{BrickRed}\xi^2}\bigg]{\color{blue}C^2_FC_A}+\bigg[-\frac{13}{36}+\frac{\zeta_3}{3}\bigg]{\color{blue} C^2_FT_Fn_f} \\ \nonumber
&+&\bigg[\frac{13}{9}+\frac{3}{8}\zeta_3-\frac{5}{6}\zeta_5-\frac{1}{48}{\color{BrickRed}\xi}-\frac{1}{96}{\color{BrickRed}\xi^2}
\bigg]{\color{blue} C_FC_AT_Fn_f}-\frac{5}{24}{\color{blue} C_FT^2_Fn^2_f}+\bigg[-\frac{20585}{9216} \\ \nonumber
&-&\frac{117}{512}\zeta_3
+\frac{55}{24}\zeta_5 
+\bigg(\frac{215}{3072}+\frac{33}{256}\zeta_3\bigg){\color{BrickRed}\xi}+\bigg(\frac{349}{3072}-\frac{9}{512}\zeta_3\bigg){\color{BrickRed}\xi^2}+\frac{9}{1024}{\color{BrickRed}\xi^3}\bigg] {\color{blue} C_FC^2_A}~,
\end{eqnarray}
\begin{eqnarray}
\label{c4mMOMBj}
c^{NS}_4&=&{\color{blue} \frac{d^{abcd}_Fd^{abcd}_A}{N_c}}\bigg[-\frac{3}{16}+\frac{\zeta_3}{4}+\frac{5}{4}\zeta_5\bigg]+{\color{blue} n_f\frac{d^{abcd}_Fd^{abcd}_F}{N_c}}\bigg[\frac{13}{16}+\zeta_3-\frac{5}{2}\zeta_5\bigg] \\ \nonumber
&+&\bigg[-\frac{4823}{2048}-\frac{3}{8}\zeta_3\bigg]{\color{blue} C^4_F}
+\bigg[-\frac{13307}{18432}-\frac{971}{96}\zeta_3+\frac{1045}{48}\zeta_5+\frac{9}{1024}{\color{BrickRed}\xi}+\frac{9}{2048}{\color{BrickRed}\xi^2}
\bigg]{\color{blue} C^3_FC_A} \\ \nonumber
&+&\bigg[\frac{2543485}{221184}+\frac{90169}{6144}\zeta_3-\frac{1375}{144}\zeta_5-\frac{385}{16}\zeta_7+\bigg(-\frac{1339}{12288}+\frac{121}{1024}\zeta_3\bigg){\color{BrickRed}\xi} \\ \nonumber
&+&\bigg(-\frac{1117}{6144}+\frac{415}{2048}\zeta_3\bigg){\color{BrickRed}\xi^2}-\frac{21}{4096}{\color{BrickRed}\xi^3}+\frac{21}{8192}{\color{BrickRed}\xi^4}
\bigg]{\color{blue} C^2_FC^2_A}
+\bigg[-\frac{3927799}{442368}+\frac{49763}{73728}\zeta_3 \\ \nonumber
&+&\frac{345755}{147456}\zeta_5+\frac{385}{64}\zeta_7-\frac{121}{96}\zeta^2_3+\bigg(\frac{107569}{147456}+\frac{1623}{2048}\zeta_3-\frac{4405}{4096}\zeta_5\bigg){\color{BrickRed}\xi} \\ \nonumber
&+&\bigg(\frac{28303}{49152}-\frac{11}{512}\zeta_3-\frac{3695}{8192}\zeta_5\bigg){\color{BrickRed}\xi^2}+\bigg(\frac{151}{3072}-\frac{59}{2048}\zeta_3-\frac{5}{4096}\zeta_5\bigg){\color{BrickRed}\xi^3} \\ \nonumber
&+&\bigg(-\frac{41}{49152}+\frac{5}{8192}\zeta_3+\frac{35}{16384}\zeta_5\bigg){\color{BrickRed}\xi^4}
\bigg]{\color{blue} C_FC^3_A}+
\bigg[\frac{317}{144}+\frac{109}{24}\zeta_3-\frac{95}{12}\zeta_5\bigg]{\color{blue}C^3_FT_Fn_f}\\ \nonumber
&+&\bigg[-\frac{6229}{864}-\frac{1739}{288}\zeta_3+\frac{205}{72}\zeta_5+\frac{35}{4}\zeta_7 
+\bigg(\frac{13}{96}-\frac{\zeta_3}{8}\bigg){\color{BrickRed}\xi}+\bigg(\frac{13}{192}-\frac{\zeta_3}{16}\bigg){\color{BrickRed}\xi^2}
\bigg]{\color{blue} C^2_FC_AT_Fn_f} \\ \nonumber
&+&
\bigg[\frac{12265}{1728}-\frac{1237}{512}\zeta_3+\frac{15}{16}\zeta_5 
-\frac{35}{16}\zeta_7
+\frac{11}{12}\zeta^2_3+\bigg(-\frac{8257}{18432}-\frac{49}{96}\zeta_3+\frac{5}{16}\zeta_5\bigg){\color{BrickRed}\xi}\\ \nonumber
&+&\bigg(-\frac{869}{3072}-\frac{33}{512}\zeta_3+\frac{5}{32}\zeta_5\bigg){\color{BrickRed}\xi^2} 
-\frac{1}{1536}{\color{BrickRed}\xi^3}+\frac{1}{3072}{\color{BrickRed}\xi^4}
\bigg]{\color{blue} C_FC^2_AT_Fn_f} \\ \nonumber
&+&\bigg[-\frac{1283}{864}+\frac{85}{72}\zeta_3-\frac{35}{36}\zeta_5-\frac{\zeta^2_3}{6}
+\bigg(\frac{11}{192}+\frac{\zeta_3}{12}\bigg){\color{BrickRed}\xi}+\frac{5}{128}{\color{BrickRed}\xi^2}
\bigg]{\color{blue} C_FC_AT^2_Fn^2_f} \\ \nonumber
&+& 
\bigg[\frac{1891}{3456}-\frac{\zeta_3}{36}\bigg]{\color{blue} C^2_FT^2_Fn^2_f}  +  \frac{5}{72}  {\color{blue} C_FT^3_Fn^3_f}~.
\end{eqnarray}
Here in expression (\ref{c4mMOMBj}) $d^{abcd}_F={\rm{Tr}}(t^at^{(b}t^ct^{d)})/6$ and $d^{abcd}_A={\rm{Tr}}(C^aC^{(b}C^cC^{d)})/6$ are the total symmetric tensors with generators $t^a$ of the fundamental representation and 
  the adjoint representation $C^a$ of the Lie algebra of the $SU(N_c)$ group.

 To  study the energy behavior of the obtained by us 
four-loop PT  expression for the 
 Bjorken polarized sum rule in the $\rm{mMOM}$ scheme   
it is necessary to take into account the gauge-dependent expression of 
the RG $\beta$-function in this scheme at the 
four-loop level, analytically evaluated in \cite{Gracey:2013sca}, and 
to define the corresponding QCD scale parameter $\Lambda_{{\rm{mMOM}}}$. 
Using  the analytical expressions for the next-to-leading order coefficients of $C^{NS}_{Bjp}$ in the $\rm{\overline{MS}}$ and $\rm{mMOM}$ schemes and 
the concept of the corresponding  effective scale parameter 
\cite{Krasnikov:1981rp}, \cite{Grunberg:1980ja},     
we obtain the following   
gauge- and flavour-dependent relation: 
\begin{equation}
\Lambda_{{\rm{mMOM}}}(\xi,  n_f)=\Lambda_{{\rm{\overline{MS}}}}\cdot{\rm{exp}}\bigg(\frac{(169+18\xi+9\xi^2)C_A-80T_Fn_f}{264C_A-96T_Fn_f}\bigg)
\end{equation}

It will be shown in  the next Section  
that  the 
values of the gauge parameters 
$\xi=-3, -1$ are highlighted by the  
presence of 
factorization of the RG $\beta$ function  in $\rm{mMOM}$ scheme variant of 
the  $\mathcal{O}(a^3_s)$ approximation of 
the  fundamental  Generalized Crewther relation for the product of 
non-singlet Adler and Bjorken functions,    
while in  the Landau gauge 
$\xi=0$ this property is true at $\mathcal{O}(a^4_s)$ order.  
In view of this it is  interesting  to consider  the asymptotic 
behavior of the PT series for the obtained by us  $\mathcal{O}(a^4_s)$
approximations of the flavour non-singlet Bjorken polarized sum rule 
in $\rm{mMOM}$ in these three theoretically prominent gauges with the 
well-known  $\rm{\overline{MS}}$ scheme results. This will be done in 
Table 1. 

\begin{center}
{\def\arraystretch{2}\tabcolsep=0.1pt
\begin{tabular}{|c|c|c|}
\hline 
$n_f$ & $\xi$  & $\textbf{$\;\;\;$The flavour NS Bjorken function} ~ \bf{C^{NS}_{Bjp}}~ \textbf{in} ~ \bf{{{\rm{\overline{MS}}}}}~ 
\textbf{and} ~ \bf{{\rm{mMOM}}}~ \textbf{schemes} $\;\;\;$ $ \\ \hline
\cline{2-3}
\multirow{4}{*}{$\;$ 3 $\;$} & $\;$ --- $\;$ &  $1-\overline{a}_s-3.583\;\overline{a}^2_s-20.2153\;\overline{a}^3_s-175.74950\;\overline{a}^4_s$   \\ \cline{2-3}
 & 0 & $1-a_s-0.896\;a^2_s+1.4262\;a^3_s-22.96225\; a^4_s$  \\
\cline{2-3}
& -1 & $1-a_s-1.083\;a^2_s+0.2312\;a^3_s-31.54404\; a^4_s$  \\
\cline{2-3}
& -3 & $1-a_s-0.333\;a^2_s-1.0317\;a^3_s-44.09174\;a^4_s$  \\
\hline
\cline{2-3}
\multirow{4}{*}{4} & --- & $1-\overline{a}_s-3.250\;\overline{a}^2_s-13.8503\;\overline{a}^3_s-102.40202\;\overline{a}^4_s$  \\
\cline{2-3}
& 0 & $1-a_s-0.840\;a^2_s+3.0375\;a^3_s-12.34185\; a^4_s$  \\
\cline{2-3}
& -1 & $1-a_s-1.028\;a^2_s+1.8633\;a^3_s-18.49192\;a^4_s$  \\
\cline{2-3}
& -3 & $1-a_s-0.278\;a^2_s+0.5170\;a^3_s-31.54819\;a^4_s$  \\
\hline
\cline{2-3}
\multirow{4}{*}{5} & --- & $1-\overline{a}_s-2.917\;\overline{a}^2_s-7.8402\;\overline{a}^3_s-41.95977\;\overline{a}^4_s$  \\
\cline{2-3}
& 0 & $1-a_s-0.785\;a^2_s+4.5099\;a^3_s-3.61660\; a^4_s$   \\
\cline{2-3}
& -1 & $1-a_s-0.972\;a^2_s+3.3566\;a^3_s-7.57175\;a^4_s$  \\
\cline{2-3}
& -3 & $1-a_s-0.222\;a^2_s+1.9269\;a^3_s-21.14145\;a^4_s$  \\
\hline
\end{tabular} }
\vspace{0.2cm}\\
{Table 1.
 The  behaviour of the PT series for $C^{NS}_{Bjp}$  at the  
$\mathcal{O}(a^4_s)$ 
level for $SU(3)$ QCD \\ with $n_f=3,4,5$  active flavours in the  
$\rm{\overline{MS}}$  and  $\rm{mMOM}$ schemes for \\ three values of the 
gauge parameter $\xi=0, -1, -3$.}
\end{center}

These results demonstrate that for all three used  gauges   
the numerical values of the gauge-dependent  coefficients, related to the  
$\rm{mMOM}$ scheme,  are considerably smaller than the ones, obtained  
in $\rm{\overline{MS}}$ scheme. Therefore, 
the convergence of the PT  series is much better in $\rm{mMOM}$ scheme.
Note also that unlike $\rm{\overline{MS}}$ $\mathcal{O}(\overline{a}^4_s)$ 
expressions for Bjorken function, the series calculated in $\rm{mMOM}$ 
scheme as a rule do not obey  sign constant behavior  character. 
It is also interesting to note, that on the contrary to the  
absolute  values of the  
$\mathcal{O}(\overline{a}^3_s)$ coefficients 
the  $\mathcal{O}(\overline{a}^4_s)$  are decreasing both in the 
$\rm{\overline{MS}}$ and  $\rm{mMOM}$-schemes with increasing number 
of quarks flavors from $n_f$=3 to $n_f=4,5$. The similiar feature is 
manifesting  itself in the process of applications of   $\rm{mMOM}$-scheme 
to the analysis of  $\mathcal{O}(a^4_s)$-approximations to other 
physical quantities (see e.g  \cite{Gracey:2014pba},
\cite{Kataev:2015yha}, \cite{Herzog:2017dtz}).   
These  facts became  noticeable starting from  the  three--loop level.

It may be of interest to compare   these numerical results with the 
ones obtained   in the process of recent analysis   
of the  PT approximations  for the 
Bjorken polarized sum rule   \cite{Ayala:2017uzx} where the $\rm{mMOM}$ was 
also used.

\section{The generalized Crewther relation in $\bf{\rm{mMOM}}$ scheme}

One of the most well-known manifestations of the consequences of the  
conformal symmetry in massless quark-parton model 
is the existence of the Crewther relation \cite{Crewther:1972kn}:
\begin{equation}
\label{Crewther}
D^{NS}C^{NS}_{Bjp}\bigg\vert_{c-i\; \rm{limit}}=\mathbbm{1}~.
\end{equation}
The unity in the l.h.s of this equation
corresponds to the normalized massless quark--parton result, obtained   
from the application of the OPE 
approach to the one-loop axial-vector-vector (AVV) triangle diagram, which defines
$\pi^0\rightarrow \gamma\gamma$ decay 
amplitude. In equation (\ref{Crewther}) the physical quantity $D^{NS}$ 
denotes the Born approximation of the   
Adler function that characterizes the 
process of one-photon  electron--positron annihilation into hadrons 
and $C_{Bjp}^{NS}$ is the normalized massless Born approximation of the 
theoretical expression for the  
Bjorken polarized sum rule, defined in (\ref{Bjp}) and (\ref{BjrNS-SI}).
However, in realistic QCD the conformal symmetry is broken. This 
leads to the existence of the  discovered in  \cite{Broadhurst:1993ru} 
generalized Crewther relation, which in $\rm{\overline{MS}}$ scheme  
can be written down as:   
\begin{equation}
\label{GCR}
D^{NS}(\overline{a}_s)C_{Bjp}^{NS}(\overline{a}_s)=\mathbbm{1}+\Delta_{csb}(\overline{a}_s) ~.
\end{equation} 
The term $\Delta_{csb}(\overline{a}_s)$ in  
(\ref{GCR}) is appearing starting from the second order of PT.   
In the related to dimensional 
regularization  
 \textit{gauge--invariant} $\rm{\overline{MS}}$   scheme, which is  commonly   used 
in the multiloop QCD calculations, its expression was first  
written down in \cite{Broadhurst:1993ru} at the $\mathcal{O}(\overline{a}_s^3)$ level 
in the following factorized form  
\begin{equation}
\label{BK}
\Delta_{csb}=\bigg(\frac{\beta(\overline{a}_s)}{\overline{a}_s}\bigg)\sum_{i\geqslant 1} K_{i}\overline{a}_s^{\; i}
\end{equation}
and confirmed at the $\mathcal{O}(\overline{a}_s^4)$ order in 
\cite{Baikov:2010je}. Here 
coefficient functions $K_i$ depend on monomials of  $SU(N_c)$ 
group structures $C_F$, $C_A$ and  $T_Fn_f$.   
The RG $\beta(\overline{a}_s)$-function is defined as
\begin{equation}  
\label{RGbeta}
\beta(\overline{a}_s)=\mu^2\frac{d\overline{a}_s}{d\mu^2}=-\sum_{i\geqslant 0}\beta_{i}{\overline{a}}^{\; i+2}_s~,
\end{equation}
In this work we will need to know its three-loop 
approximation only, which was  computed in the $\rm{\overline{MS}}$ 
scheme in \cite{Tarasov:1980au} and confirmed later on in 
\cite{Larin:1993tp}.
 The corresponding expression of the  $\beta$ function in 
the $\rm{mMOM}$ scheme can be found in analytical form in \cite{Gracey:2013sca}, \cite{Ruijl:2017eht} and depends on $\xi$ beginning from two--loop level.

It is important now to understand whether the fundamental 
property of factorization of the conformal anomaly term $(\beta(\overline{a}_s)/\overline{a}_s)$ in 
(\ref{BK}) is fulfilled in the \textit{gauge--invariant} $\rm{\overline{MS}}$--like schemes only.  
To study  this problem we will extend the consideration  of the representation (\ref{BK}) 
to the \textit{gauge--dependent} $\rm{mMOM}$ scheme.

Regardless of the used  renormalization schemes  the physical quantities $D^{NS}$ and $C^{NS}_{Bjp}$
in (\ref{GCR}) obey the property of the renormalization invariance. 
Using this property and omitting the details of considerations 
we obtain the following transition  relations for  the    $K_i$ terms in (\ref{BK}) from
 $\rm{\overline{MS}}$ scheme to any other renormalization scheme AS: 
\begin{eqnarray}
\label{K1-2}
K^{AS}_1&=&K^{{\rm{\overline{MS}}}}_1~, ~~~~~~~
K^{AS}_2=K^{{\rm{\overline{MS}}}}_2+\bigg(\frac{\beta^{{\rm{\overline{MS}}}}_1-\beta^{AS}_1}{\beta_0}+2\delta^{AS}_1\bigg)K^{{\rm{\overline{MS}}}}_1~, \\
\label{K3}
K^{AS}_3&=&K^{{\rm{\overline{MS}}}}_3+\bigg(\frac{\beta^{{\rm{\overline{MS}}}}_1-\beta^{AS}_1}{\beta_0}+3\delta^{AS}_1\bigg)K^{{\rm{\overline{MS}}}}_2+ \\ \nonumber
&+&\bigg(\frac{\beta^{{\rm{\overline{MS}}}}_2-\beta^{AS}_2}{\beta_0}+\frac{ 3\beta^{{\rm{\overline{MS}}}}_1 -2\beta^{AS}_1 }{\beta_0}\delta^{AS}_1-\frac{\beta^{{\rm{\overline{MS}}}}_1-\beta^{AS}_1}{\beta_0}\cdot\frac{\beta^{AS}_1}{\beta_0}+2\delta^{AS}_2+(\delta^{AS}_1)^2\bigg)K^{{\rm{\overline{MS}}}}_1
\end{eqnarray}
We  emphasize that  if all the fractions included in these expressions are explicitly divided by 
QCD $\beta_0$--factors, then  the property of factorization of the conformal symmetry breaking 
factor $(\beta(a^{AS}_s)/a^{AS}_s)$   in AS scheme
will be valid at the $\mathcal{O}(a^{AS\; 3}_S)$ at least.  The  terms $\delta^{AS}_i$ in (\ref{K1-2}), (\ref{K3})  are the AS--analogies of the $\delta_i$ terms in (\ref{aMS-mMOM}). If we fix $\rm{mMOM}$ scheme instead of AS scheme, we obtain that at $\mathcal{O}(a^2_s)$ level the factorization of $\beta(a_s)/a_s$ function for $K_2$ term  is possible for three distinguished values of the gauge parameter $\xi$  only,  namely for $\xi=0$ (Landau gauge), for $\xi=-1$  (anti--Feynman gauge) and for $\xi=-3$ \cite{Stefanis:1983ke}, \cite{Mikhailov:1998xi} (Stefanis--Mikhailov gauge).  At the   $\mathcal{O}(a^3_s)$  approximation the $\beta(a_s)/a_s$ factorization property is valid  for the  
Landau gauge only  and the partial factorization holds  for anti--Feynman and Stefanis--Mikhailov gauges (when the factorization condition is imposed, then for one of the six possible color monomials the concrete  coefficient is not determined).   
Thus  we conclude that total  factorization of the conformal anomaly  $\beta(a_s)/a_s$  in the GCR is also possible for \textit{gauge--dependent} renormalization schemes. Therefore we rule out the gauge invariance as a cause of the factorization property. Theoretical reasons of these 
our findings are yet  unclear to us. The detailed description of the outlined in this talk our studies are in progress.

\section{Acknowledgments}
The authours are grateful to S.V. Mikhailov for  constructive remarks. 
One of us (A.K) would like to thank organizers of  XVIIth International Workshop on High Energy Spin  Physics (DSPIN-17) and A.V. Efremov personally  
for hospitality in  Dubna.  
His  work   on studies  of the theoretical features of the  generalized Crewther relation in QCD  is supported by  the Russian Science 
Foundation Grant No. 14-22-00161. 
The work of V.M related to the calculation and analysis of the non-singlet Bjorken function in $\rm{mMOM}$ scheme is 
supported by  the Russian Science Foundation grant No. 16-12-10151.

\end{document}